\documentclass[12pt]{article}
\usepackage{graphicx}

\def\pbnr{}
\def\speaker{F.Fernandez, D. R. Entem, P.G. Ortega}
\def\title{Molecular Structures in Hidden Charm Meson and Charmed Baryon Spectrum}
\def\affiliation{Nuclear Physiscs Group and IUFFyM\\
The University of Salamanca, Salamanca, Spain}

\newcommand\pubnumber{\pbnr}
\newcommand\pubdate{\today}
\def\Title#1{\begin{center} {\Large #1 } \end{center}}
\def\Author#1{\begin{center}{ \sc #1} \end{center}}

\newcommand{\OnBehalf}[1]{\sbox0{#1}\ifdim\wd0=0pt
        {}
	\else
	{\\on behalf of #1}
	\fi}
\newcommand{\SupportedBy}[1]{\sbox0{#1}\ifdim\wd0=0pt
        {}
	\else
	{\footnote{#1}}
	\fi}
\def\Address#1{\begin{center}{ \it #1} \end{center}}

\newcommand\pubblock{\includegraphics[width=5cm]{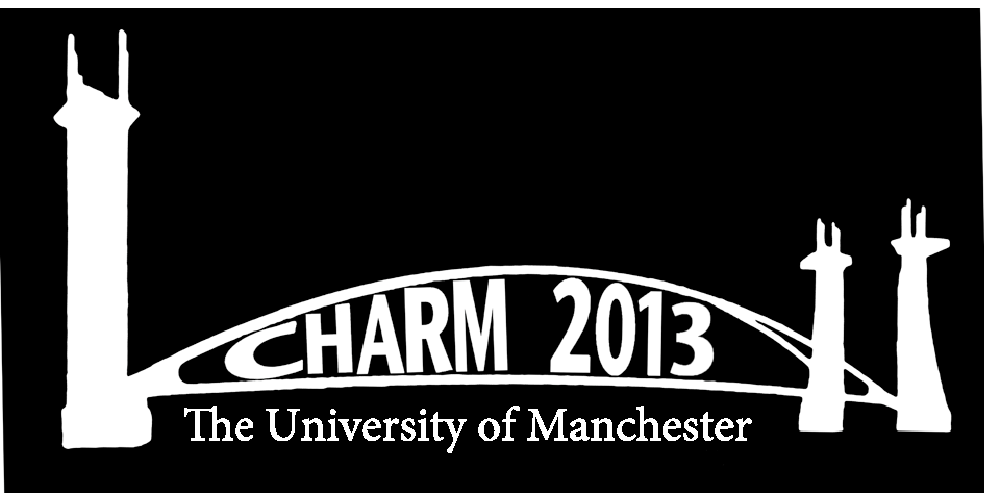}\hfill{\begin{tabular}{l} \pubnumber\\
         \pubdate  \end{tabular}}}
\newenvironment{Abstract}{\begin{quotation}  }{\end{quotation}}
\newenvironment{Presented}{\begin{quotation} \begin{center} 
             PRESENTED AT\end{center}\bigskip 
      \begin{center}\begin{large}}{\end{large}\end{center} \end{quotation}}
\def\Acknowledgements{\bigskip  \bigskip \begin{center} \begin{large}
             \bf ACKNOWLEDGEMENTS \end{large}\end{center}}
\def\venue{The 6$^{th}$ International Workshop on Charm Physics\\
(CHARM 2013)\\
Manchester, UK,  31 August -- 4 September, 2013}

\def\beq{\begin{equation}}
\def\eeq#1{\label{#1}\end{equation}}
\def\eeqn{\end{equation}}

\def\beqa{\begin{eqnarray}}
\def\eeqa#1{\label{#1}\end{eqnarray}}
\def\eeqan{\end{eqnarray}}

\let\bar=\overbar

\def\Dslash{\not{\hbox{\kern-4pt $D$}}}
\def\dslash{\not{\hbox{\kern-2pt $\del$}}}

\def\msb{{\bar{\ssstyle M \kern -1pt S}}}

\begin{document}
\begin{titlepage}
\pubblock

\vfill
\Title{\title}
\vfill
\Author{\speaker}
\Address{\affiliation}
\vfill
\begin{Abstract}
Using a constituent quark model we study the mass and decay channels of meson meson and meson baryon structures
in the charm sector.

We show that the $X(3872)$ and $X(3940)$ resonances can be described as mixed charmonium-molecular states
with $J^{PC}=1^{++}$, whereas the $X(3915)$ and the $Y(3940)$ can be assigned to similar 
mixed states with $J^{PC}=0^{++}$.

In the baryon spectrum we identify the $\Lambda^+_c(2940)$ as a $D^*N$ molecule with $(I)J^P=(0)3/2^-$ and the recently reported $X_c(3250)$ as a $D^*\Delta$ resonance with $(I)J^P=(1)5/2^-$ or $(I)J^P=(2)3/2^-$.
\end{Abstract}
\vfill
\begin{Presented}
\venue
\end{Presented}
\vfill
\end{titlepage}
\def\thefootnote{\fnsymbol{footnote}}
\setcounter{footnote}{0}
%

\section{Introduction}

In the last several years many exotic hadronic states has been observed
with unexpected properties that disfavor their interpretation as ordinary 
hadrons.

Most of these states appear near open flavor thresholds and therefore one can expect
that loosely bound or resonant meson-meson or baryon-meson structures mix with the original $q\bar q$ or
$qqq$ structures to configure the physical states.

The $X(3872)$ was the first of this kind of states to be discovered. Its decays into $J/\psi \rho$
and $J/\psi \omega$ channels violate isospin and completely rules out a $c\bar c$ interpretation.
Recently LHCb has fixed its quantum numbers~\cite{LHCb1}. Moreover other candidate to this quantum 
numbers appear in this mass region, namely the $X(3940)$, which was observed  by Belle~\cite{Ab07}.

Two years after the discovery of the $X(3872)$, the $Y(3940)$ was observed by the Belle Collaboration~\cite{Choi05}. Belle reported a mass of $M=3943\pm 11\pm 13$ MeV and a width $\Gamma =87\pm 22\pm 26$ MeV.  A later measurement from BABAR~\cite{Amo10} confirmed this resonance but with a mass value $M=3919\pm 3.8\pm 2.04$ MeV. Later on, a new  state, the $X(3915)$, has been reported by the Belle Collaboration in the $\gamma\gamma\to J/\psi\omega$ decay~\cite{ueh10}. The measured mass is $M=3914\pm3\pm2$ MeV and the width $\Gamma=23\pm9$ MeV.
The proximity of the masses could indicate that these states are connected to the same particle~\cite{nora}. However there are evidences that the two states could be molecular states. The $Y(3940)$ has a lower limit for the decay $J/\psi \omega$ of $\Gamma > 1$ MeV, which is large for an OZI suppressed channel of conventional charmonium. On the other hand the product $\Gamma_{\gamma \gamma}\times B[X(3915)\rightarrow J/\psi \omega]$ is relatively large compared to charmonium predictions. Very recently, the quantum numbers of the $X(3915)$ were established as $J^\pi=0^+$~\cite{Lees}.

These situations also appear in the baryon spectrum, where the $\Lambda_c^+(2940)$, observed by Babar~\cite{BB1} and later confirmed by Belle~\cite{BLL1}, is just 6 MeV/$c^2$ below the $D^*p$ threshold. Similarly the recently reported 
$X_c(3250)$~\cite{BB2} state is close to the $D^*\Delta$ threshold. 

In this work we propose a unified theoretical explanation of these states as meson meson or meson baryon  molecular states in a constituent quark model that has been extensively used to describe the hadron phenomenology~\cite{US1,US3}.

The model is based on the assumption that the spontaneous chiral symmetry breaking  generates the constituent mass. To compensate this mass term in the hamiltonian the Lagrangian must include Goldstone boson fields which mediates the interaction between quarks. The minimal realization of this mechanism include a pseudo scalar boson and an scalar one.
In this model baryons (mesons) are described as clusters of three quark (one pair of quark and antiquark). This fact does not affect the heavy quark sector but is of paramount importance in the molecular picture because the only remaining  interaction between the two molecular component, due to its color singlet nature, is the one driven by the Golsdtone boson exchanges between the light quarks. 

The meson meson and the meson baryon interactions are obtained using the Resonanting Group method. The coupling between the $q\bar q$ and the four quarks configuration are performed using the $^3P_0$ model. A more datailed discussion of the model and its application to the states mentioned above can be found in~\cite{JPG}.

\section{Results for the charmonium sector}

We have performed a calculation including the two quark $c\bar c (2^3P_1)$ together with the $D^0D^{*0}$, $D^{\pm}D^{*\mp}$, $J/\psi \rho$ and $J/\psi \omega$ channels. The coupling of the $DD^*$ with the $J/\psi \rho$ and $J/\psi \omega$ channels is not enough to bind the molecule and it is mandatory to couple the $c\bar c$ pair to reach a molecular bound state.
To introduce the isospin breaking in our calculation we will work in the charge basis instead of the isospin symmetric one, allowing the dynamics of the system to choose the weight of the different components. Isospin is explicitly broken using the experimental meson masses.

The different components for the $X(3872)$ an the $X(3940)$ are shown in Table~\ref{t6}.

\begin{table}[!t]
\begin{center}
\begin{tabular}{cccccc}
\hline
\hline
 $Mass$ & $c\bar c(2 ^3P_1)$ & $D^0{D^*}^0$ & $D^
\pm{D^*}^\mp$ & 
$J/\psi\rho$ & $J/\psi\omega$ \\
\hline
   $3871.5$ & $8.00$  & $86.61$  & $4.58$ & $0.53$ & $0.29$ \\
   $3941.8$ & $61.09$  & $18.53$  & $16.85$ & $0.01$ & $3.52$ \\ 
\hline
\hline
\end{tabular}
\caption{\label{t6} Binding energy (in MeV) and
channel probabilities  (in $\%$)
for the $X(3872)$ and $X(3940)$ states.}
\end{center}
\end{table}

The reults for the experimental ratios $R_1=\frac{X(3872)\rightarrow \pi^+\pi^-\pi^0J/\psi}{X(3872)\rightarrow 
\pi^+\pi^-J/\psi}=0.8\pm 0.3$ and $R_2=\frac{\Gamma(X(3872)\to\gamma\Psi(2S))}{\Gamma(X(3872)\to\gamma J/\psi)}\leq 2.1 ( 90\% C.L.)$  are respectively $0.44$ and $1.2$ in agreement with the data. 

As seen in Table~\ref{t6}, the $X(3940)$ has almost the same value of neutral and charged $DD^*$ components. Therefore the isospin breaking has almost completely dissapear and the resonance is basically $I=0$. This justifies that the coupling with the $J/\psi \rho$ channel is almost negligible. 

The $X(3872)$ decays basically through the $DD^*$ component being the branching ratio for the different channels $\mathcal{B.R.}(X(3872)\to DD^*)=0.89$, $\mathcal{B.R.}(X(3872)\to J/\psi \omega)=0.1$, $\mathcal{B.R.}(X(3872)\to J/\psi \rho)=3\times10^{-4}$.
 In this way the puzzle between this two states with the same quantum numbers seems to be solved in a satisfactory way.

We have performed the same calculation as before but for the $J^{PC}=0^{++}$ sector. We include the $2^3P_0$ $c\bar c$ channel together with the molecular channels $D\bar D$(3736.05 MeV), $J/\psi \omega$(3879.56 MeV), $D_s\bar D_s$(3936.97 MeV), $J/\psi \phi $(4116.0 MeV) where the channel thresholds are indicated between brackets.  
The results are shown in Table~\ref{t7}.

We find a narrow state which can be identified with the $X(3915)$. This resonance is basically a mixture of $c\bar c$ and $D\bar D$ molecular components. Moreover a second  wide resonance appears at $M=3970$ MeV. This resonance can be the old $Y(3940)$, today dissapeared from the PDG. Its dominant component is the $c\bar c(2^3P_0)$ although the contribution of the molecular  $D\bar D$ is also significant. This structure provide a possible explanation for the unusual decay mode $J/\psi \omega$ through the rescattering $0^{++}\rightarrow D\bar D\rightarrow J/\psi \omega$.

\begin{table}[!t]
\begin{center}
\begin{tabular}{cccccccc}
\hline
\hline
 $ state$ & $Mass$ & $\Gamma$ &  $c\bar c(2^3P_0)$ & $D^0 \bar D^0$ & $J/\psi\omega$ & $D_s\bar D_s$ & $J/\psi\phi$ \\
\hline
   $X(3915)$   & $3896.5$ & $4.10$ & $34.22$  & $46.67$  & $9.42$ & $9.67$ & $0.03$ \\ 
   $Y(3940)$   & $3970$ & $189.3$ & $57.27$  & $35.32$  & $0.15$ & $5.72$ & $1.54$ \\ 
\hline
\hline
\end{tabular}
\caption{\label{t7} Mass and total width (in MeV) and
channel probabilities  (in $\%$)
for the $X(3915)$ and $Y(3940)$.}
\end{center}
\end{table}

\section{The $\Lambda^+_c(2940)$ and the $X_c(3250)$}

To explain the $\Lambda^+_c(2940)$ we have studied the possible molecular sytructures of the $ND^*$ system. We found a molecular state with $J^P=\frac{3}{2}^-$ quantum numbers at a mass of $2938,68$ MeV/$c^2$. One can see from Table~\ref{t1}  that the molecule is basically a $^4S_{3/2}$ $D^*N$ state with a small mixture of  $D_{3/2}$ states. 

\begin{table}
\begin{center}
\begin{tabular}{c|ccc|cc|cc}
\hline
\hline
$M\,(MeV)$ & $\mathcal P_{^4S_{3/2}}$ & $\mathcal P_{^2D_{3/2}}$& $\mathcal P_{^4D_{3/2}}$
& $\mathcal P_{{D^*}^0p}$ & $\mathcal P_{{D^*}^+n}$&$\mathcal P_{I=0}$ & $\mathcal P_{I=1}$ \\
\hline
   2938,80& 96,22 & 0,86 & 2,92& 63,93  & 36,07 & 97,52 & 2,48 \\
\hline
\hline
\end{tabular}
\caption{\label{t1} Mass of the $\Lambda_c^+(2940)$ and contributions for the different channels.}
\end{center}
\end{table}

\begin{table}
\begin{center}
\begin{tabular}{c|c|c|c}
\hline
\hline
Decay channel & $Width\,(MeV)$ & decay channel& $Width\,(keV)$ \\
\hline
$\Lambda_c^+ \rightarrow D^0p$ & 9.42 & $\Lambda_c^+\rightarrow \Sigma_c^{++} \pi^-$ & 29.7 \\
$ \Lambda_c^+ \rightarrow D^+n$ & 10.74 & $\Lambda_c^+\rightarrow \Sigma_c^{+} \pi^0$ & 25.2 \\
&  & $\Lambda_c^+\rightarrow \Sigma_c^0 \pi^+$ & 21.1 \\
\hline
\hline  
 $\Gamma (total)$ & 20.2 & $\Gamma (experimental)$ & $17^{+8}_{-6}$ \\
\hline
\hline
\end{tabular}
\caption{\label{t2} Widths of the $\Lambda_c^+(2940)$ for different decay channels.}
\end{center}
\end{table}

The different components in the charge basis show that the $\Lambda_c^+(2940)$ is  a $D^{*0}p$ molecule with a sizable  $D^{*+}n$ component. Looking to the isospin one  realizes that the state is almost a pure $I=0$ state as required by the experimental data.
The results for the different decay widths are shown in Table~\ref{t2}. 
The  widths of the $D^0p$ and $D^+n$ decay channels are of the order of the experimental width whereas
the contributions of the $\Sigma\pi$ channels are negligible. The predicted total width also agree with the experimental data.

\begin{table}
\begin{center}
\begin{tabular}{cccc}
\hline
\hline
$J^P$ & $ I $ & $Mass (Mev/c^2)$ & $P_{max}(Channel)$ \\
\hline
   $\frac{1}{2}^-$ & 2 & 3232,7 & 99,71($^2S_{1/2}$)  \\
   $\frac{3}{2}^-$ & 2 & 3238,2 & 99,69($^4S_{3/2}$)  \\
   $\frac{5}{2}^-$ & 1 & 3226,1 & 97,25($^6S_{5/2}$)  \\
\hline
\hline
\end{tabular}
\caption{\label{t3} Quantum numbers and masses of the different $D^*\Delta$ resonances}
\end{center}
\end{table}

Finally, in the $D^*\Delta$ sector, there are three possible  states in the mass range of $3230$ MeV/$c^2$ (see Table~\ref{t3}). They are basically $L=0$ resonances.

All these states can decay through the $\Sigma_c^{++} \pi^-\pi^-$ channel by rearrangement. Nevertheless this is not the main decay channel for the $D^*\Delta$ molecule because it should be easier to decay through the $D^*N\pi$ channel. This is the channel where the experimentalist should look to confirm the structure of the $X_c(3250)$.

\section {Summary}

As a summary, we have found that molecular structures may play an important role in the description of the meson and baryon spectra. In particular the $X(3872)$, $X(3940)$, $X(3915)$ and $Y(3940)$ can be describes as $DD^*$ resonances coupled to $q\bar q$ states. Without changing any parameter we found a $D^*N$ molecule in the $J^P=\frac{3}{2}^-$ channel and $I=0$ with a mass and total width that agrees with the experimental data for the $\Lambda_c^+(2940)$ baryon. Finally we found that the recently reported $X_c(3250)$ can be explained as a $D^*\Delta$ molecule.

\Acknowledgements
This work has been partially funded by Ministerio de Ciencia y Tecnolog\'\i a
under Contract
No. FPA2010-21750-C02-02, 
by the European Community-Research Infrastructure Integrating
Activity ``Study of Strongly Interacting Matter'' (HadronPhysics3 
Grant no. 283286) and
by the Spanish Ingenio-Consolider 2010 Program
CPAN (CSD2007-00042).

\end{document}